# Nanothermomechanics

ANDRZEJ TRZĘSOWSKI

**Institute of Fundamental Technological Research**
**Polish Academy of Sciences**
Pawińskiego 5B, 02-106 Warsaw, Poland
e-mail: atrzes@ippt.gov.pl or atrzes@wp.pl

The paper concerns the dependence of thermomechanical properties of three-dimensional solid nanoclusters on the cluster size as well as on its shape. Investigations are restricted to the class of so-called homogeneous thermodynamic processes with kinematics based on affine group and referred to the one whole body, not an infinite system of subbodies. It is shown that then the thermodynamics of nanoclusters is consistent with dynamics of affinely-rigid bodies (constrained or not and elastic as well as admitting viscosity effects). The main topics discussed are: (i) a group-theoretical description of structurally stable solid nanoclusters; (ii) a phenomenological model of mechanical properties of nanoclusters revealing the coexistence of solid and liquid states in a finite interval of absolute temperature.

## 1.  Introduction

Between the dimensions of an atomic scale and the dimensions which characterize bulk materials is to be found a size range where condensed matter exhibits some remarkable specific properties [1]. One particular phenomenon – the dependence of a cluster physical properties upon its size and shape – occurs for clusters in the nanometer scale. For example, the strength of *nanoclusters* (i.e. compact three-dimensional aggregates of atoms and/or molecules with the mean size not greater than 100 nm) increases when the cluster size decreases [2]. The elastic moduli of such clusters reveal also the size effect [3]. These are examples of *mechanical size effects*.

It is also observed that metallic nanoclusters manifest the *size-dependent structural transitions* of their material structure. Namely, not only the crystal structure of small metallic particles can be dependent on their size (e.g. No, Mo, W and Ta particles of diameters 5-10 nm have face-centered cubic or hexagonal structure in place of the usual volume-centered cubic lattice), but even some such particles can lose their crystal structure and become amorphous [4].

The shape of a nanostructure can depend on its size. Namely, it is observed that sodium clusters with a small number of atoms (<150-200) crystallize in the form of icosahedra. The structure becomes unstable for a large number of atoms and transforms to cubo-octahedra (i.e. a cube with truncated corners), which is just a path of the face-centered cubic lattice [5, 6]. More generally, when a large number of metallic atoms aggregate in a slow manner at low temperature, then they take the shape o a regular polyhedron with the close-packed structure [5]. Moreover, it is known that the structurally most stable are such symmetric crystalline nanoclusters that are invariant under the action of the point symmetry group of their crystal structure [7]. It is a *shape effect*.

There exist also the *thermodynamic size effects*. For example, the reduction of the melting point of small gold aggregates as a function of decreasing particle size is observed [1, 4]. Note that although it is only possible to formulate a valid theoretical description of the melting point in the thermodynamic limit (i.e. for those systems whose dimensions are infinite), nevertheless, it is possible to make an experimental determination of the melting temperature of a small system. Three different criteria could be used for this determination [1]: (i) the disappearance of the state of order in the solid; (ii) the sharp variation of some physical properties: evaporation rate, magnetic susceptibility, etc.; (iii) the sudden change in the particle shape (e.g. it is attributed to the transition from polyhedral to spherical shapes, undergone by some crystals to an effect associated with melting of the solid [1]).

More generally, we are dealing with the problem of a comprehensive understanding of the thermodynamics of finite (and macroscopically small) material systems that can contain e.g. even less than 200 atoms [6, 8]. Consequently, such systems can not be satisfactorily described, in a manner consistent with the phenomenological classical thermodynamics, by means of the classical statistical physics. However, for example, a finite system of mutually interacting identical particles, the dynamical behavior of which is random, can admit the thermodynamical interpretation of the Markovian-type evolution of the system ([9, Part II] and [10]). Namely, there can be shown the existence of such thermodynamically permitted Markov-type processes in small systems that are consistent with the assumption of thermal character of the interaction of the system with the environment, with the first and second laws of thermodynamics, with the postulate of the existence of a stationary (in general nonequilibrium), uniquely defined Gibbs-type state, and with the relaxation postulate stating that the process relax, independently of the choice of the initial condition, towards to this Gibbs-type state. Moreover, if the environment of the system is a thermostat, then we can generalize, identifying the uniform temperature attributed to the system with the (uniform) temperature of the thermostat, the equilibrium definition of the free energy to the nonequilibrium situation [10]. This example shows that the classical thermodynamics can be consistent with the randomness of microstate dynamics of macroscopically small systems containing a small number of atoms or molecules.

If nanoclusters are treated as *macroscopically small continuous bodies*, then their deformation and temperature can be treated as those approximated to uniform state variables of the cluster. Particularly, it means that nanoclusters can be considered as macroscopically small *"affinely-rigid"* bodies (i.e. homogeneously deformed bodies with translational degree of freedom of their mass center), with dynamics being a generalization of the rigid body dynamics and originally formulated as a theory which is not associated with an observation scale [11]. The uniform temperature of a thermostat can be attributed, according to the previous remarks, to the nanocluster and interpreted as a measure of its thermal state. Note that this definition of the temperature of a small system is consistent with the above-mentioned method of a experimental determination of its melting temperature. Thus, owing to these approximations, we are dealing with the so-called *homogeneous thermodynamic processes* [12, 13] with kinematics based on the affine group and considered as being dependent on the body size and shape. The thermomechanical properties of nanoclusters can be described then in the framework of such classical thermodynamics which refers to one whole system, not an infinite family of subsystems ([9, 16] and Sections 2-5). It is a phenomenological version of the so-called *nanothermodynamics* (e.g. [14] and [15]), extended by a non-local model of dynamics of isolated nanoclusters ([9, 16] and Sections 2, 6 and 7).

If the mean size of clusters is not greater than 100 nm, then the dependence of thermomechanical properties of these clusters on their size as well as their shape, becomes physically significant effects (it is especially visible in the case of polycrystalline nanoclusters with the size grains less than 10 nm) [2, 17]. That is why we restrict ourselves to the case of such nanoclusters only. Note also that the proposed approach to the description of nanoclusters can be applied, for example, to the investigations of "classical" behavior of the molecules of the fullerene $C_{60}$. This particular case is interesting from the point of view of the question if and how the quantum theory applies to macroscopic objects. $C_{60}$ is of course not a macroscopic object, but these molecules have a mass at least one order of magnitude greater than that of any other object whose wave properties have been previously observed [18]. It extends the applicability of wave-particle duality by about one order of magnitude in the macroscopic direction. However, it should be stressed that there are about 15 orders of magnitude to go before we reach the mass of anything we would normally think of as macroscopic [18].

## 2. Thermoelastic nanoclusters

We will deal with homogeneous, compact and connected nanoclusters (called also size-effect bodies [14, Part I]) of immovable center of mass, homogeneously deformed end endowed with an uniform temperature (see Section 1). Spatial configurations of such a *body* can be identified with subsets $B$ of the three-dimensional Euclidean point space $E^3$ with a distinguished point $o \in E^3$ (say e.g. – the center of mass of the nanocluster). The point space $E^3$ can be identified then with the Euclidean vector space $\mathrm{E}^3$ of translations in $E^3$, the spatial configurations of the body can be considered as subsets $B \subset \mathrm{E}^3$ of the form $B = l(\mathbf{F})(B_0)$, where $B_0 \subset \mathrm{E}^3$ is a distinguished spatial configuration of the body called its *reference configuration* and identified with the body itself, $l(\mathbf{F})$ denotes the following linear mapping in $\mathrm{E}^3$:

$$\mathbf{x} = l(\mathbf{F})(\mathbf{X}) = \mathbf{F}\mathbf{X}, \quad \mathbf{X} \in B_0, \quad \mathbf{F} \in GL^+(\mathrm{E}^3),$$

(2.1)

$$GL^+(\mathrm{E}^3) = \{\mathbf{F} \in L(\mathrm{E}^3): \det \mathbf{F} > 0\}, \quad L(\mathrm{E}^3) = \mathrm{E}^3 \otimes \mathrm{E}^3,$$

and the tensor space $L(\mathrm{E}^3)$ is the 9-dimensional Euclidean vector space with respect to the standard scalar product defined by $\mathbf{A} \cdot \mathbf{B} = \mathrm{tr}(\mathbf{A}\mathbf{B}^\mathrm{T})$ for $\mathbf{A}, \mathbf{B} \in L(\mathrm{E}^3)$, and $\mathbf{B}^\mathrm{T}$ denotes the transposition of $\mathbf{B}$. The mass $m$ of the body $B_0$ is the same for all its spatial configurations $B$, and the volumetric mass density $\rho$ of these configurations is defined by:

(2.2)
$$m = \rho_0 \mathrm{V}(B_0) = \rho \mathrm{V}(B),$$
$$\mathrm{V}(B) = J(\mathbf{F}) \mathrm{V}(B_0), \quad J(\mathbf{F}) = \det \mathbf{F},$$

where $\mathrm{V}(B)$ denotes the volume of $B$. Note that if particles constituting a cluster $B_0$ are close-packed and have the same mass density $\rho_p$, then the mass density $\rho_0$ of this body is approximated by $\rho_p$. It is e.g. the case of fullerene $C_{60}$ clusters [19] but it is not the case of single $C_{60}$ molecules with all carbon atoms located in vertices of a truncated icosahedron.

The domain of the deformation gradient $\mathbf{F}$ may be restricted, for physical reasons (cf. the size effect discussed Section 1), to an open subset $M \subset GL^+(\mathrm{E}^3) \subset L(\mathrm{E}^3)$. More generally, we may consider *mechanically constrained nanoclusters* defined by the condition that the set $M$ of admissible deformation gradients of a nanocluster is a connected differential manifold such that [20]

(M1) $M \subset GL^+(\mathrm{E}^3)$,

(M2) $\mathbf{1} \in M$,

(M3) $\mathbf{Q} \in SO(\mathrm{E}^3), \mathbf{F} \in M \Rightarrow \mathbf{QF} \in M$,

where $\mathbf{1}$ denotes the unity element of the Lie group $GL^+(\mathrm{E}^3)$ and $SO(\mathrm{E}^3)$ denotes the proper orthogonal group on $\mathrm{E}^3$. The condition (M2) states that a reference configuration $B_0$ of the body is consistent with constraints and the condition (M3) means that all rigid motions of the body are admitted by constraints. We will say that $B_0$ is an *unconstrained* body if $\dim M = 9 (= \dim GL^+(\mathrm{E}^3))$ and that $B_0$ has con-

*straints* if $\dim M < 9$. Particularly, we will say that the body has *scalar constraints* if [13, 20]

(2.3) $$M = \{\mathbf{F} \in GL^+(\mathrm{E}^3) : h(\mathbf{F}) = 0\},$$

where $h \in C^k(U)$, $k \geq 2$, $M \subset U$ and $U$ is an open and connected subset of the Lie group $GL^+(\mathrm{E}^3)$. The condition (M3) enables to define $h$ as the so-called *objective scalar*; that is, for each $\mathbf{F} \in M$, the following condition should be fulfilled [13]:

(2.4) $$\forall \mathbf{Q} \in SO(\mathrm{E}^3), \quad h(\mathbf{QF}) = h(\mathbf{F}).$$

Let $I \subset \mathbb{R}_+$ be an interval of absolute temperatures. In the classical thermodynamics, the mechanical influence on a body $B_0$ can be introduced into the theory by what is called *Gibbs form* on $M \times I$ defined by the formula

(2.5) $$\Omega = dE - \theta dS + \mathbf{N} \cdot d\mathbf{F},$$

where $\theta \in I$ is an absolute temperature of the body $B_0$ (see remarks in Section 1), $d\mathbf{F}$ is an infinitesimal deformation gradient increment along an arbitrary direction in $L(\mathrm{E}^3)$ and the tensor field $\mathbf{N}$ with values in $L(\mathrm{E}^3)$ is called the *generalized thermodynamic force*. The scalars $E$ and $S$ of class $C^k$, $k \geq 2$, are defined for all *thermodynamic configurations* $\lambda = (\mathbf{F}, \theta) \in M \times I$ and denote the total *internal energy* and the *entropy* of the body $B_0$, respectively. These scalars are related to the total *Helmholtz free energy* $\Psi$ of the body by the so-called Legendre transformation:

(2.6) $$\Psi(B_0; \lambda) = E(B_0; \lambda) - \theta S(B_0; \lambda).$$

We will assume that the scalars $E$, $\Psi$ and $S$ are extensible to scalars of class $C^k$, $k \geq 2$, defined on an open and connected set $U$ such that $M \subset U \subset GL^+(\mathrm{E}^3)$. The dependence of these scalars on the (compact and connected) geometrical figure $B_0$ represents the dependence of thermomechanical properties of the nanocluster upon its size and shape (Section 1). Moreover, these scalars would be *objective* according to the rule (2.4) (with $B_0$ and $\theta$ treated as parameters). The influence of mechanical action (i.e. the generalized thermodynamic force $\mathbf{N}$) on the change of internal energy and entropy is described by the following *dissipation inequality*:

(2.7) $$\Omega(B_0; \lambda) \leq 0,$$

where $\lambda = (\mathbf{F}, \theta) \in M \times I$. The nanocluster is called then *thermoelastic*.

An *unconstrained* thermoelastic nanocluster does not reveal dissipation because then, for each thermodynamic configuration $\lambda = (\mathbf{F}, \theta) \in GL^+(E^3) \times I$, we have [13]:

(2.8) $$\Omega(B_0; \lambda) = 0.$$

Eq.(2.8) is equivalent to the following formulae defining the *thermomechanical response* of $B_0$:

(2.9) $$\mathbf{N} = -\frac{\partial \Psi}{\partial \mathbf{F}}, \quad S = -\frac{\partial \Psi}{\partial \theta},$$

and stating that the free energy function is the so-called thermodynamic potential. A *stationary state* of this nanocluster $B_0$ is defined as the thermodynamic configuration $\lambda_0 = (\mathbf{F}_0, \theta_0) \in GL^+(E^3) \times I$ such that

(2.10) $$d\Psi(B_0; \lambda_0) = 0,$$

what is equivalent to the following conditions:

(2.11) $$\mathbf{N}(B_0; \lambda_0) = \mathbf{0}, \quad S(B_0; \lambda_0) = 0.$$

The stationary state of the form $\lambda_0 = (\mathbf{1}, \theta_0)$ will be called *natural* at the temperature $\theta_0 \in I$.

Let $B_0$ be a (mechanically) *constrained* nanocluster endowed with the scalars $E$, $\Psi$ and $S$ of Eqs.(2.5)-(2.7), extensible to scalars of class $C^k$, $k \geq 2$, defined on an open and connected set $U$ such that $M \subset U \subset GL^+(E^3)$. Note that then, even in the "*purely mechanical*" case of isothermal ($d\theta = 0$) and isentropic ($dS = 0$) states for which the dissipation inequality reduces to the following condition:

(2.12) $$\Omega = d\Psi + \mathbf{N} \cdot d\mathbf{F} \leq 0,$$

Eq.(2.8) with $\lambda \in M \times I$ is, in general, incorrect [20]. Consequently, the free energy function is not here a thermodynamic potential. In this paper we postulate that the mechanical constraints are *ideal* in this sense that Eq.(2.8) and the condition

(2.13) $$S(B_0; \lambda) = -\partial_\theta \Psi(B_0; \lambda),$$

are fulfilled for each thermodynamic configuration $\lambda \in M \times I$. In this case, the generalized thermodynamic force $\mathbf{N}$ can be written in the form

(2.14) $$\mathbf{N}(B_0; \lambda) = -\partial_\mathbf{F} \Psi(B_0; \lambda) + \mathbf{N}_c(B_0; \lambda),$$

where $\mathbf{N}_c$ is an undefined tensor field on $M \times I$ such that

(2.15) $\qquad \boldsymbol{\lambda} = (\mathbf{F}, \theta) \in M \times I \;\Rightarrow\; \mathbf{N}_c(B_0; \boldsymbol{\lambda}) \in \mathrm{T}_\mathbf{F}^\perp M,$

and $\mathrm{T}_\mathbf{F}^\perp M \subset L(\mathrm{E}^3)$ is the orthogonal complement in the Euclidean tensor space $L(\mathrm{E}^3)$ of the vector space $\mathrm{T}_\mathbf{F} M \subset L(\mathrm{E}^3)$, tangent to $M$ at the point $\mathbf{F} \in M$; that is, independently of the choice of the temperature $\theta \in I$, we have:

(2.16) $\qquad \forall \dot{\mathbf{F}} \in \mathrm{T}_\mathbf{F} M, \; \mathbf{N}_c(B_0; \mathbf{F}, \theta) \cdot \dot{\mathbf{F}} = 0.$

Particularly, for scalar constraints we have (cf. [13]):

(2.17) $\qquad \mathbf{N}_c(B_0; \mathbf{F}, \theta) = \alpha(B_0; \mathbf{F}, \theta) \partial_\mathbf{F} h(\mathbf{F}),$

where $\alpha$ is an undefined scalar.

The *objectivity condition* of a generalized thermodynamic force $\mathbf{N}$ takes, for each thermodynamic configuration $\boldsymbol{\lambda} = (\mathbf{F}, \theta) \in M \times I$, the following form [13]:

(2.18) $\qquad \forall \mathbf{Q} \in SO(\mathrm{E}^3), \; \mathbf{N}(B_0; \mathbf{QF}, \theta) = \mathbf{Q}\mathbf{N}(B_0; \mathbf{F}, \theta).$

It follows from the objectivity of the Helmholtz free energy function $\Psi$ and from Eq.(2.14) that Eq.(2.18) reduces, for mechanically constrained bodies, to the objectivity condition of the undefined generalized constraint force $\mathbf{N}_c$. Particularly, for scalar constraints this reduces itself, according to Eq.(2.4), to the objectivity condition of the undefined scalar $\alpha$.

The above-discussed compact and connected nanoclusters, unconstrained or ideally constrained and revealing the objective thermomechanical response, can be identified as *thermoelastic*. Note that, contrary to the thermoelastic simple materials (see e.g. [13]) for which the dissipation coming from the heat conduction appears, the such defined thermoelastic nanoclusters are *perfectly thermoelastic* in this sense that they are elastic within a certain range of temperature.

### 3. Solid nanoclusters

Let us observe that if the size and shape effects of a perfectly thermoelastic body $B_0$ (constrained or unconstrained – Section 2) can be neglected, then the tensor field $\hat{\mathbf{T}}: M \times I \to L(\mathrm{E}^3)$ of the form

$$\text{(3.1)} \qquad \hat{\mathbf{T}} = -\mathrm{V}(B_0)^{-1}\mathbf{N},$$

where the tensor field $\mathbf{N}$ is defined by Eq.(2.14) and $\mathrm{V}(B_0)$ denotes the volume of the body $B_0$ (identified with its reference configuration – Section 2), reduces to the so-called Piola stress tensor for (homogeneously deformed) simple thermoelastic bodies (see e.g. [13]). Thereby, we can consider $\hat{\mathbf{T}}$ as a *generalized Piola stress tensor* assigned to the one whole nanocluster $B_0$. Consequently, the one-parameter field of surface forces $\hat{\mathbf{t}}_\theta$, $\theta \in I$, of the form:

$$\text{(3.2)} \qquad \hat{\mathbf{t}}_\theta(B_0;\mathbf{F},\mathbf{X}) = -\hat{\mathbf{T}}(B_0;\mathbf{F},\theta)\mathbf{n}_0(\mathbf{X}),$$

where $\lambda = (\mathbf{F},\theta) \in M \times I$ and $\mathbf{n}_0(\mathbf{X})$, $\mathbf{X} \in \partial B_0$, denotes the outward normal unit vector at the point $\mathbf{X}$ of the boundary cluster $\partial B_0$, can be interpreted as a field of non-local internal forces acting on the surface $\partial B_0$. It ought to be stressed that although the tensor $\hat{\mathbf{T}}$ of Eq.(3.1) is a global counterpart of the Piola stress tensor, it is not a measure of stress as normally understood. Nevertheless, $\mathbf{N}^\mathrm{T}$ can be interpreted as a dipole moment $\hat{\mathbf{M}}_{\text{int}}$ of internal surface forces defined by $\hat{\mathbf{T}}$ according to Eq.(3.2) and acting on the nanocluster boundary. Namely, since [9, 21]:

$$\text{(3.3)} \qquad \hat{\mathbf{M}}_{\text{int}}(B_0;\mathbf{F},\theta) = \int_{\partial B_0} \mathbf{X} \otimes \hat{\mathbf{t}}_\theta(B_0;\mathbf{F},\mathbf{X})\, dF(\mathbf{X}),$$

and thus

$$\text{(3.4)} \qquad \hat{\mathbf{T}} = -\mathrm{V}(B_0)^{-1}\hat{\mathbf{M}}_{\text{int}}^\mathrm{T},$$

we can take the dipole moment as a global measure of these internal surface forces. We obtain then the following interpretation rule of the generalized thermodynamic force:

$$\text{(3.5)} \qquad \mathbf{N} = \hat{\mathbf{M}}_{\text{int}}^\mathrm{T}.$$

We will consider also the *generalized Cauchy stress tensor* defined as [9, 16]:

$$\text{(3.6)} \qquad \mathbf{T}(B_0;\mathbf{F},\theta) = J(\mathbf{F})^{-1}\hat{\mathbf{T}}(B_0;\mathbf{F},\theta)\mathbf{F}^\mathrm{T} = -\mathrm{V}(B)^{-1}\mathbf{N}(B_0;\mathbf{F},\theta)\mathbf{F}^\mathrm{T},$$

where $B = l(\mathbf{F})(B_0)$ denotes the deformed spatial configuration of $B_0$, and Eqs.(2.1), (2.2) and (3.1) were taken into account. Particularly, it follows from the objectivity of the free energy function and from Eqs.(2.9), (2.18), (3.1) and (3.6) that the generalized Cauchy stress tensor is, in the case of unconstrained perfectly ther-moelastic nanoclusters (Section 2), a symmetric and objective tensor function with respect to the variable $\mathbf{F}$ (see e.g. [13]) of the form [9, 16]:

(3.7)
$$\mathbf{T}(B_0; \mathbf{F}, \theta) = \mathbf{R}\mathbf{h}(B_0; \mathbf{U}, \theta)\mathbf{R}^T = \mathbf{T}(B_0; \mathbf{F}, \theta)^T,$$
$$\mathbf{h}(B_0; \mathbf{U}, \theta) = V(B)^{-1} \partial_{\mathbf{U}} \Psi(B_0; \mathbf{U}, \theta) \mathbf{U},$$

where the so-called polar decomposition of $\mathbf{F}$ (e.g. [13]) was taken into account:

(3.8)
$$\mathbf{F} = \mathbf{R}\mathbf{U}, \quad \mathbf{R} \in SO(\mathrm{E}^3), \quad \mathbf{U} = \mathbf{U}^T \in GL^+(\mathrm{E}^3).$$

Introducing the one-parameter field of internal forces $\mathbf{t}_\theta$, $\theta \in I$, acting on the boundary $\partial B$ of the deformed spatial configuration $B$ of $B_0$:

(3.9)
$$\mathbf{t}_\theta(B_0; \mathbf{F}, \mathbf{x}) = -\mathbf{T}(B_0; \mathbf{F}, \theta)\mathbf{n}(\mathbf{x}),$$
$$\mathbf{x} = \mathbf{F}\mathbf{X} \in \partial B, \quad \mathbf{X} \in \partial B_0,$$

where $\mathbf{n}$ is the field of outward unit vectors normal to $\partial B$, we obtain that the dipole moment $\mathbf{M}_{int}$ of these forces defined as

(3.10)
$$\mathbf{M}_{int}(B_0; \mathbf{F}, \theta) = \int_{\partial B} \mathbf{x} \otimes \mathbf{t}_\theta(B_0; \mathbf{F}, \mathbf{x}) dF(\mathbf{x}),$$

has the following representation:

(3.11)
$$\mathbf{M}_{int}(B_0; \mathbf{F}, \theta) = \mathbf{F}\hat{\mathbf{M}}_{int}(B_0; \mathbf{F}, \theta) = -V(B)\mathbf{T}(B_0; \mathbf{F}, \theta),$$

and thus, according to Eqs.(2.18) and (3.6), it is also a symmetric and objective tensor function with respect to the variable $\mathbf{F} \in M$.

Now, we can define the *insensibility group* $G_\theta(B_0)$ at the temperature $\theta \in I$ of the *unconstrained and perfectly thermoelastic* nanocluster $B_0$ as [9, 16]:

(3.12)
$$G_\theta(B_0) = \{\mathbf{H} \in SL(\mathrm{E}^3): \forall \mathbf{F} \in GL^+(\mathrm{E}^3), \mathbf{T}(B_0; \mathbf{F}\mathbf{H}, \theta) = \mathbf{T}(B_0; \mathbf{F}, \theta)\},$$
$$SL(\mathrm{E}^3) = \{\mathbf{F} \in GL^+(\mathrm{E}^3): \det \mathbf{F} = 1\}.$$

We will say, imitating the concept of simple solid materials (see [13]), that the considered unconstrained and perfectly thermoelastic nanocluster $B_0$ is, within the range $I$ of temperature, an *undistorted solid nanocluster*, if [16]

(3.13)
$$\forall \theta \in I, \; G_\theta(B_0) \subset SO(\mathrm{E}^3).$$

It can be shown that then (see [13] and [16]):

(3.14) $G_\theta(B_0) = \{\mathbf{Q} \in SO(\mathrm{E}^3): \forall \mathbf{F} \in GL^+(\mathrm{E}^3), \Psi(B_0; \mathbf{F}\mathbf{Q}, \theta) = \Psi(B_0; \mathbf{F}, \theta)\}.$

Note that in particular applications concerning solid bodies, the existence of an *unstressed* spatial configuration of the body is usually assumed [13]. We will denote by $B_0 = B_{\theta_0}$ a thermoelastic undistorted solid nanocluster such that

(3.15) $\quad\quad\quad\quad \exists \theta_0 \in I, \ \mathbf{T}_0(B_0; \mathbf{1}, \theta_0) = \mathbf{0}.$

For example, it is the case of nanocluster $B_0$ being in a natural stationary state at the temperature $\theta_0 \in I$ (Section 2).

We will also say, taking into account remarks concerning the structural stability of nanoclusters (Section 1), that the (perfectly thermoelastic) undistorted solid nanocluster is, within the range *I* of temperature, a *structurally stable solid nanocluster*, if [16]:

(3.16) $\quad\quad\quad\quad \forall \theta \in I, \ G_\theta(B_0) \subset g(B_0),$

where $g(B_0)$ is the group of rotational symmetries of the geometrical figure $B_0$:

(3.17) $\quad\quad\quad\quad g(B_0) = \{\mathbf{Q} \in SO(\mathrm{E}^3): \ l(\mathbf{Q})(B_0) = B_0\}.$

### 4. Liquid-like response of nanoclusters

The science of nanoscale concerns the properties and behavior of mesoscale aggregates of atoms and/or molecules, at a scale not yet large enough to be considered macroscopic but far beyond what can be called microscopic (Section 1). As we shrink the mesoscale to the *nanoscale*, physics becomes increasingly dominated by the surfaces. Particularly, most of the unique features of crystalline nanoclusters arise from a very high ratio of the number of surface atoms to the total number of atoms in the cluster. The well-known phenomenon of the dependence of shape of fullerenes on the number of their surface atoms is an example of such a property of nanoclusters [19]. Therefore, in the case of nanoclusters, the surface energy substantially affects the properties of the bulk material [6]. Note also that, on the nanoscale observation level, the notion of the state of matter takes a new meaning. Namely, it follows from theoretical predictions, confirmed by computer modeling and experimental observations, that at least for some *particular sizes* of the crystalline nanoclusters they would exhibit a coexistence of solid and liquid states within a finite range of temperature near the melting point. It is supposed that in the coexistence region, a nanocluster fluctuates back and forth between its solid state (with the lowest-free energy) and its liquid state (with the higher-free energy). Thus, it is the case when we can only state that, within a finite range temperature and pressure, the nanocluster occurs in the condensed state [4, 19]. This makes small clusters different from the bulk systems whose solid and liquid phases coexist only at a single temperature point, the melting point. Consequently, taking into account these properties of nano-clusters and their properties discussed in Section 1, one may even attempt to identify nanostructures as constituting a new phase of matter – the *nanomatter* [6].

The above remarks suggest to consider a class of such perfectly thermoelastic nanoclusters (Section 2) that reveal some properties of solids as well as fluids. First of all, let us observe that a fluid is commonly regarded as a material having "no preferred configurations" [13]. We will say, in agreement with this statement, that a perfectly thermoelastic nanocluster $B_0$ (Section 2) reveals the *liquid-like response* if its total free energy function $\Psi$ fulfils the following condition [9]:

$$(4.1) \quad \forall (\mathbf{F}, \theta) \in M \times I, \quad \Psi(B_0; \mathbf{F}, \theta) = \hat{\Phi}_\theta\big(l(\mathbf{F})(B_0)\big),$$
$$\hat{\Phi}_\theta : W_M(B_0) \to \mathbb{R}, \quad W_M(B_0) = \{B = l(\mathbf{F})(B_0), \ \mathbf{F} \in M\}.$$

The objectivity condition of the free energy function and the definition of mechanically constrained nanoclusters (Section 2) mean that the set $W_M(B_0)$ of all spatial configurations of $B_0$ is closed under the action of proper orthogonal group $SO(\mathrm{E}^3)$ and that the functionals $\hat{\Phi}_\theta$, $\theta \in I$, are objective; that is, for each $B \in W_M(B_0)$, the condition

$$(4.2) \quad \forall \mathbf{Q} \in SO(\mathrm{E}^3), \quad \hat{\Phi}_\theta\big(l(\mathbf{Q})(B)\big) = \hat{\Phi}_\theta(B),$$

is fulfilled. If $M = GL^+(\mathrm{E}^3)$, then $B_0$ is an unconstrained nanocluster and the set of all its spatial configurations is denoted $W(B_0)$.

Let us consider the case when the nanocluster $B_0$ as well as its certain deformed configuration $B_\mathbf{P} = l(\mathbf{P})(B_0) \in W(B_0)$, $\mathbf{P} \in GL^+(\mathrm{E}^3)$, are unconstrained and perfectly thermoelastic *undistorted solid* nanoclusters (Section 3) with the liquid-like response defined, according to Eq.(4.1) with $W_M(B_0) = W(B_0) = W(B_\mathbf{P})$, by the one-parameter family of functionals $\hat{\Phi}_\theta : W(B_0) \to \mathbb{R}$, $\theta \in I$. It follows from Eqs.(3.13), (3.14), (3.17), (4.1) and (4.2) that it should be [16]:

(4.3) $\quad\quad\quad \forall \theta \in I, \; g(B_\mathbf{P}) \subset G_\theta(B_\mathbf{P}) = \mathbf{P} G_\theta(B_0) \mathbf{P}^{-1} \subset SO(\mathrm{E}^3).$

Thus, applying the polar decomposition formula of Eq.(3.8):

(4.4) $\quad\quad\quad\quad\quad\quad \mathbf{P} = \mathbf{R}(\mathbf{P}) \mathbf{U}(\mathbf{P}),$

we obtain that [13, 16]

(4.5) $\quad\quad\quad\quad\quad\quad G(B_\mathbf{P}) = \mathbf{R}(\mathbf{P}) G_\theta(B_0) \mathbf{R}(\mathbf{P})^{-1}$

and

(4.6) $\quad\quad\quad\quad \forall \mathbf{Q} \in G_\theta(B_0), \; \mathbf{Q} \mathbf{U}(\mathbf{P}) \mathbf{Q}^\mathrm{T} = \mathbf{U}(\mathbf{P}).$

For example, if $B_\theta = l(\mathbf{P}_\theta)(B_0)$ and $B_0 = B_{\theta_0}$, $\theta, \theta_0 \in I$, are unstressed spatial configurations of the nanocluster $B_0$, that is (Section 3):

(4.7) $\quad\quad\quad\quad\quad \mathbf{T}(B_\theta; \mathbf{1}, \theta) = \mathbf{T}(B_0; \mathbf{1}, \theta_0) = \mathbf{0},$

where $\mathbf{T}$ denotes the generalized Cauchy stress tensor (Section 3), then Eqs.(4.3)-(4.6) with $\mathbf{P} = \mathbf{P}_\theta$, $\theta \in I$, describe, within a certain range of temperature $I$, the influence of temperatures on the insensibility groups of (unconstrained and perfectly thermoelastic) undistorted solid nanoclusters with the liquid-like response [16].

If the undistorted solid nanocluster $B_0$ with the liquid-like response is additionally *structurally stable* (Section 3), then, according to Eq.(3.16), Eq.(4.3) with the condition

(4.8) $\quad\quad\quad\quad\quad \forall \theta \in I, \; G_\theta(B_0) = g(B_0)$

should be taken into account. Note that, according to the well-known theorem (e.g.[22], [23]), there exists a correspondence between finite subgroups of the proper rotation group $SO(\mathrm{E}^3)$ and the symmetries of *regular polyhedrons*. Namely, symmetries of a *tetrahedron* define the tetrahedron group *T*, symmetries of an *octahedron* define the octahedron group *O* and symmetries of an *icosahedron* define the icosahedron group *I*. A *cube* has the same symmetry group as an octahedron and a *dodecahedron* has the same symmetry group as an icosahedron [23]. Particularly, it follows from Eq.(4.6) that if

(4.9) $\quad\quad\quad\quad\quad\quad\quad g(B_0) = O,$

then

(4.10) $$\mathbf{U}(\mathbf{P}) = \alpha \mathbf{1}, \quad \alpha > 0.$$

Moreover, according to Eqs.(3.17), (4.5), (4.6) and (4.8), Eq.(4.10) is also valid for

(4.11) $$g(B_0) = SO(\mathrm{E}^3),$$

and $B_0$ and $B_\mathbf{P}$ being *balls* of a radius $R_0$ and $R = \alpha R_0$, respectively. In this case

(4.12) $$\forall \theta \in I, \quad G_\theta(B_\mathbf{P}) = g(B_\mathbf{P}) = SO(\mathrm{E}^3).$$

If

(4.13) $$g(B_0) = T \subset O,$$

then

(4.14) $$\mathbf{U}(\mathbf{P}) = \alpha \mathbf{1} + \beta \mathbf{k} \otimes \mathbf{k},$$
$$\alpha > 0, \quad \alpha + \beta > 0, \quad \mathbf{k} \cdot \mathbf{k} = 1,$$

and Eq.(4.14) is also valid if

(4.15) $$g(B_0) = G(\mathbf{k}),$$

is the group of all rotations about an axis parallel to the unit vector $\mathbf{k}$. In this case

(4.16) $$\forall \theta \in I, \quad G_\theta(B_\mathbf{P}) = g(B_\mathbf{P}) = G(\mathbf{R}(\mathbf{P})\mathbf{k}) = \mathbf{R}(\mathbf{P}) G(\mathbf{k}) \mathbf{R}(\mathbf{P})^{-1},$$

and the boundaries $\partial B_0$ (of $B_0$) and $\partial B_\mathbf{P}$ (of $B_\mathbf{P}$) would be e.g. *surfaces of revolutions* with the axes of revolution parallel to $\mathbf{k}$ and $\mathbf{R}(\mathbf{P})\mathbf{k}$, respectively.

The groups $SO(\mathrm{E}^3)$, $G(\mathbf{k})$ and the icosahedron group $I$ are not crystallographic point groups [13, 22, 23]. However, the groups $O$ and $T$ are point groups corresponding to the *cubic* and to the *tetragonal* (and *hexagonal*) crystallographic systems, respectively [13, 22, 23]. Therefore, structurally stable are e.g. the undistorted perfectly thermoelastic nanoclusters $B_0$ and $B_\mathbf{P} = l(\mathbf{P})(B_0)$, where $\mathbf{P} \in GL^+(\mathrm{E}^3)$ is defined by Eqs.(4.4) and (4.6), such that: (i) $B_0$ is an isotropic ball or a ball with a cubic crystal lattice; (ii) $B_0$ has the boundary $\partial B_0$ being a surface of revolution and this nanocluster is transversally isotropic or with a tetragonal or hexagonal crystal lattice. Let us observe that in both above examples, the corresponding tensors $\mathbf{U}(\mathbf{P})$ (of Eq.(4.10) in the case (i) or of Eq.(4.14) in the case (ii)) define *shape-preserving deformations* $B_0 \to B_\mathbf{P}$. For example, if tensors $\mathbf{P} = \mathbf{P}_\theta$, $\theta \in I$, are defined by Eq.(4.7) and addi-tionally $\mathbf{R}(\mathbf{P}_\theta) = \mathbf{1}$, $\theta \in I$, then these shape-preserving deformations of the nanocluster $B_0$ can be identified with *free thermal distortions* of this nanocluster and thus the family $B_I = \{B_\theta, \theta \in I\}$ can be recognized as the one consisting of *thermally equivalent nanoclusters*.

Note that, as it was mentioned in Section 1, metallic nanoclusters with very small number of atoms can crystallize (in contrast to their bulk crystalline counterparts) in the form of *icosahedra*. It is in line with our concept of liquid-like response of (macroscopically small) solid nanoclusters because an icosahedral short-range arrangement has been found in liquids [24].

## 5. Quasi-solid state

The notion of liquid-like response of nanoclusters introduced in Section 4 is consistent with the group-theoretical description of mechanical properties of solids as well as fluids, formulated in the framework of the theory of simple materials (e.g. [13]). Let us consider, in order to illustrate this consistency, the one-parameter family of functionals $\hat{\Phi}_\theta : W \to \mathbb{R}$, $\theta \in I$, where $W$ is the set of all compact and *convex* three-dimensional bodies in $E^3$ endowed with the so-called Hausdorff (or Blashe) metric [25, 26], such that for each $B_0 \in W$ the following condition holds (cf. (4.1)):

(5.1) $\quad \forall (\mathbf{F}, \theta) \in GL^+(E^3) \times I, \quad \Psi(B_0; \mathbf{F}, \theta) = \hat{\Phi}_\theta (l(\mathbf{F})(B_0))$.

It can be shown (basing on the Hadwiger integral theorem – [25-27]) that if these functionals are continuous, invariant with respect to the action of isometry group in $E^3$ and additive (in this sense that $\hat{\Phi}_\theta(B_1 \cup B_2) = \hat{\Phi}_\theta(B_1) + \hat{\Phi}_\theta(B_2) - \hat{\Phi}_\theta(B_1 \cap B_2)$ for every $B_1, B_2 \in W$)), then they can be represented in the following general form:

(5.2) $\quad \hat{\Phi}_\theta(B) = \mathrm{a}(\theta)\mathrm{V}(B) + \mathrm{b}(\theta)\mathrm{F}(B) + \mathrm{c}(\theta)\mathrm{M}(B) + \mathrm{d}(\theta)$,

where $\mathrm{a}(\theta)$, $\mathrm{b}(\theta)$, $\mathrm{c}(\theta)$, $\mathrm{d}(\theta)$ are arbitrary constants assumed here to be functions of class $C^2$ of the temperature parameter $\theta \in I$. $\mathrm{V}(B)$, $\mathrm{F}(B)$ and $\mathrm{M}(B)$ denote the volume of the domain $B$, the surface field of its boundary $\partial B$ and the total mean curvature of $\partial B$, respectively. The quantity $\mathrm{M}(B)/2\pi$, known in rock analysis and in stereographic metallography, is interpreted as the mean grain width [25, 27]. Eqs.(5.1) and (5.2) admit the case of convex bodies with a piecewise smooth boundaries, i.e. containing some edges and corners [26, 27]. For example, if $B_0$ is a cube with edges of length $l_0$ directed along $\mathbf{e}_k$-directions of an orthonormal base $\{\mathbf{e}_k;\ k=1, 2, 3\}$ in the Euclidean vector space $\mathrm{E}^3$, $B = l(\mathbf{F})(B_0)$, where the tensor $\mathbf{F} \in GL^+(\mathrm{E}^3)$ is defined by Eq.(3.8) and the following condition:

(5.3)
$$\mathbf{U}\mathbf{e}_k = \lambda_k \mathbf{e}_k, \quad k = 1, 2, 3,$$
$$\mathbf{e}_k \cdot \mathbf{e}_l = \delta_{kl}, \quad \lambda_k > 0,$$

then $B$ is a rectangular parallelepiped whose concurrent edges have directions $\mathbf{a}_k = \mathbf{R}\mathbf{e}_k$, lengths $l_k = \lambda_k l_0$, $k = 1, 2, 3$, and the following formulae hold [16, 30]:

(5.4)
$$\mathrm{V}(B) = l_1 l_2 l_3 = l_0^3 \mathrm{III},$$
$$\mathrm{F}(B) = 2(l_1 l_2 + l_2 l_3 + l_3 l_1) = 2l_0^2 \mathrm{II},$$
$$\mathrm{M}(B) = \pi(l_1 + l_2 + l_3) = \pi l_0 \mathrm{I},$$

where I, II and III are the well-known *principal invariants* of the deformation tensor $\mathbf{F}$, which are used in the theory of isotropic elastic materials [13]:

(5.5)
$$\mathrm{I} = \mathrm{tr}\mathbf{U} = \lambda_1 + \lambda_2 + \lambda_3,$$
$$\mathrm{II} = \frac{1}{2}\left[(\mathrm{tr}\mathbf{U})^2 - \mathrm{tr}\mathbf{U}^2\right] = \lambda_1\lambda_2 + \lambda_2\lambda_3 + \lambda_3\lambda_1,$$
$$\mathrm{III} = \det\mathbf{U} = \lambda_1\lambda_2\lambda_3.$$

Conversely, for any deformation tensor $\mathbf{F} \in GL^+(\mathrm{E}^3)$ we can define, according to Eqs.(3.8) and (5.3), such a cube $B_0$ that principal invariants of $\mathbf{F}$ can be represented by the geometric parameters of Eq. (5.4). This example suggests that the functionals V, F, M : $W \to \mathbb{R}_+$ of Eq.(5.2) can be proposed as global counterparts of principal invariants of the deformation tensors considered in the local theory of elastic materials.

Note that although elastic fluids are such isotropic elastic materials for which stresses are dependent on the principal invariant III of Eq.(5.5) only [13], the formula

(5.6)
$$\hat{\Phi}_\theta(B) = \varepsilon(\theta)\mathrm{V}(B) + \gamma(\theta)\mathrm{F}(B),$$
$$\varepsilon(\theta) \geq 0, \quad \gamma(\theta) > 0,$$

is considered in the classical capillarity theory. The constant $\varepsilon(\theta)$ depends on bulk interatomic interactions only and thus it vanishes in the case of incompressible fluids (or solids). The constant $\gamma(\theta)$ is the so-called *surface tension* and it is a quantity attributed to the boundary thin film (infinitely thin, i.e. reduced to the boundary surface, in the phenomenological approximation used here). Since the surface tension can be identified with the free energy density needed to change the boundary surface field unit, even the boundary solid surface can be endowed with this quantity [28]. In this case, $\gamma(\theta)$ can be considered as a quantity conditioned by the interactions of atoms located on the boundary solid surface [28, 29] and it is a positive quantity at the temperature lower than the melting temperature [28]. In the case of solids bodies for which their size is much greater than an effective size of the boundary layer, the influence of the mean curvature on the total free energy $\Psi$ of Eqs.(5.1) and (5.2) can be neglected and thus the formula (5.6) can be then accepted [29]. However, it is not the case of nanoclusters for which the formula (5.2) with

(5.7) $\qquad \varepsilon(\theta) = |a(\theta)| \geq 0, \quad \gamma(\theta) = |b(\theta)| > 0, \quad \omega(\theta) = |c(\theta)| \geq 0,$

ought to be taken into account. The quantity $2\pi\omega(\theta)$ can be interpreted then as the free energy density needed to change the nanocluster mean width $M(B)/2\pi$ unit. On the other hand, the mean curvature is a relative geometric quantity depending on the Euclidean geometry of the ambient physical space in which the boundary surface is embedded. Consequently, the quantity $\omega(\theta)$ should be considered as the one conditioned by interactions between the boundary surface atoms and bulk atoms located in a boundary layer [9].

Finally, it seems physically reasonable to consider a non-local nanoscale counterpart of simple thermoelastic isotropic materials (solids as well as liquids – [13]) defined as perfectly thermoelastic, compact and convex nanoclusters with the liquid-like response of the form given by Eq.(5.1) and by the following condition [16]:

(5.8) $\qquad \forall B \in W, \quad \hat{\Phi}_\theta(B) = \Phi_\theta(V(B), F(B), M(B)),$

where $\Phi_\theta : \mathbb{R}^3_+ \to \mathbb{R}$, $\mathbb{R}^3_+ = \{x = (x^1, x^2, x^3) \in \mathbb{R}^3 : x^k \geq 0, \ k = 1, 2, 3\}$, $\theta \in I$, are mappings of class $C^2$. The generalized Cauchy stress tensor is given then by [16]:

$$\mathbf{T}(B_0; \mathbf{F}, \theta) = p_\theta(\mathrm{V}, \mathrm{F}, \mathrm{M})\mathbf{1} + \mathbf{R}\mathbf{t}(B_0; \mathbf{U}, \theta)\mathbf{R}^\mathrm{T},$$

(5.9)
$$\mathbf{t}(B_0; \mathbf{U}, \theta) = \frac{1}{\mathrm{V}}\left(\frac{\partial \Phi_\theta}{\partial \mathrm{F}}\partial_\mathbf{U}\mathrm{F} + \frac{\partial \Phi_\theta}{\partial \mathrm{M}}\partial_\mathbf{U}\mathrm{M}\right)\mathbf{U},$$

$$p_\theta(\mathrm{V}, \mathrm{F}, \mathrm{M}) = \frac{\partial \Phi_\theta}{\partial \mathrm{V}}(\mathrm{V}, \mathrm{F}, \mathrm{M}),$$

where Eqs.(3.7), (3.8), (5.1) and (5.8) were taken into account. Thus, according to Eqs.(3.12) and (3.17), we have

(5.10) $$\forall \theta \in I, \quad g(B_0) \subset G_\theta(B_0) \subset SL(\mathrm{E}^3).$$

Note that, according to Eqs.(3.12) and (5.8)-(5.10), the condition

(5.11) $$\forall \theta \in I, \quad G_\theta(B_0) = SL(\mathrm{E}^3),$$

is equivalent to the following formula:

(5.12) $$\forall (B, \theta) \in W \times I, \quad \hat{\Phi}_\theta(B) = \Phi_\theta(\mathrm{V}(B)).$$

It follows from Eq.(5.9) that the condition (5.11) describes an *ideal fluid* or *gas*. So, if Eq.(5.8) does not reduce to Eq.(5.12), then the condition (3.13) can be assumed. The nanocluster $B_0 \in W$ is then an undistorted and perfectly thermoelastic *solid nanocluster* such that

(5.13) $$\forall \theta \in I, \quad g(B_0) \subset G_\theta(B_0) \subset SO(\mathrm{E}^3),$$

and thus, according to Eq.(3.16), this nanocluster is *structurally stable* iff the conditon (4.8) is fulfilled.

We see that the definition of the liquid-like response of a *convex nanocluster* $B_0$ given by Eqs.(5.1), (5.8) and (5.9) admits the case of its solid state as well as gaseous or liquids states. Moreover, it follows from Eq.(5.9) that while the scalar $p$ depends on the actual spatial configuration of the nanocluster only, the tensor $\mathbf{t}$ depends on the choice of its reference configuration. We will say, taking into account these statements and remarks at the beginning of the Section 4, that the free energy function of Eqs.(5.1) and (5.8) defines the *quasi-solid state* of a perfectly thermoelastic convex nanocluster $B_0$. The case of Eq.(5.13) concerns an *undistorted solid nanocluster* being in the quasi-solid state [16].

It has been observed that the most stable small metallic clusters have an *almost spherical shape*. The oblate or prolate shape of such a cluster means that its structure is less stable [5]. Let us define, in order to introduce a measure of degree of sphericity of nanoclusters being in quasi-solid state, the following effective radii:

$$r_V = \left(\frac{3V}{4\pi}\right)^{1/3}, \quad r_F = \left(\frac{F}{4\pi}\right)^{1/2}, \quad r_M = \frac{M}{4\pi}, \tag{5.14}$$

where V, F and M denote the volume of a nanocluster $B \in W$ (being in an undeformed or deformed state), the surface field and the total mean curvature of its boundary, respectively. The quantity

$$d = 2r_M, \tag{5.15}$$

can be interpreted as the *mean width* of the nanocluster (see remarks following Eq.(5.2)). It is known that the following inequalities hold:

$$r_V \leq r_F \leq r_M. \tag{5.16}$$

In each of these three relations equality is attained in the case of a ball, and in this case only [31]. Moreover, introducing the new variables $\xi$ and $\eta$ by:

$$\xi = \frac{r_F}{r_M}, \quad \eta = \frac{r_V}{r_M}, \tag{5.17}$$

we obtain that

$$0 < \eta \leq \xi^{4/3} \leq 1, \quad 0 < \kappa = \eta/\xi = r_V/r_F \leq 1, \tag{5.18}$$

and $\xi = 1$ or $\eta = 1$ or $\kappa = 1$ iff the nanocluster is a ball [31]. Therefore, the variables $\xi$, $\eta$ and $\kappa$ define equivalent *shape coefficients* measuring the degree of sphericity of nanoclusters under consideration. In the literature is also considered the following shape coefficient K [32] equivalent to $\kappa$:

$$K = \frac{F_V V}{V_F F} = \kappa^5,$$

$$F_V = 4\pi r_V^2, \quad V_F = (4/3)\pi r_F^3. \tag{5.19}$$

For example (cf. Section 4), an *icosahedron* is the regular polyhedron of the highest shape coefficient ($K = 0.835$ - [32]). It is the case of metallic nanoclusters with a very small number of atoms (Section 1). Consequently, we can expect that the nanoclusters of the shape of regular polyhedra, observed for a large number of atoms (Section 1), correspond to less stable shapes of nanoclusters than the icosahedral nanocluster is. Namely, we have [32]: $K = 0.791$ for a dodecahedron, $K = 0.657$ for octahedron, $K = 0.583$ for a cube, and $K = 0.370$ for a tetrahedron.

It seems also physically reasonable to restrict ourselves to the functions $\Phi_\theta$ of Eq.(5.8) symmetric with respect to their arguments V, F, M. So, let us assume that

(5.20)
$$\Phi_\theta(\mathrm{V},\ \mathrm{F},\ \mathrm{M}) = \varphi_\theta(s_1,\ s_2,\ s_3),$$
$$s_n = s_n(r_\mathrm{V},\ r_\mathrm{F},\ r_\mathrm{M}), \quad n = 1,\ 2,\ 3,$$

where $s_1$, $s_2$ and $s_3$ are three functionally independent, fully symmetric functions of the effective radii. For example, each symmetric polynomial with respect to the variables $r_\mathrm{V}$, $r_\mathrm{F}$ and $r_\mathrm{M}$ can be expressed as a polynomial with respect the following *elementary symmetric functions* [33]:

(5.21)
$$s_n = r_\mathrm{V}^n + r_\mathrm{F}^n + r_\mathrm{M}^n = \left(\frac{d}{2}\right)^n \left(1 + \xi^n + \eta^n\right), \quad n = 1,\ 2,\ 3,$$

where $d$ is the mean nanocluster width defined by Eq.(5.15). For physical reasons (see Section 1), the following condition should be taken into account:

(5.22)
$$\frac{d}{l_0} \in \langle \alpha,\ \beta \rangle,$$

where as $l_0$ can be taken a characteristic length of the nanocluster (e.g. $l_0 = d_0$ - the mean nanocluster width for which it exhibits a coexistence of solid and liquid states). Note that if $l_0$ is the critical length associated with a certain physical property of clusters, then the case $\beta < 1$ can describe the phenomenon that this property changes for the nanocluster mean with $d$ fulfilling the condition (5.22) (see e.g. [2] and [34]).

## 6. Thermodynamic processes

Let us consider a *homogeneous process* [13] defined as a curve of thermodynamic configurations (Section 2):

(6.1)
$$\lambda = (\mathbf{F},\ \theta): T \to M \times I, \quad T \subset \mathbb{R},$$

where $T$ is a time interval. Then

(6.2)
$$\dot{\lambda}(t) = \frac{d}{dt}\lambda(t) = \left(\dot{\mathbf{F}}(t),\ \dot{\theta}(t)\right) \in \mathrm{T}_{\mathbf{F}(t)}M \times \mathbb{R},$$
$$\mathrm{T}_{\mathbf{F}(t)}M \subset L(\mathrm{E}^3), \quad \dim \mathrm{T}_{\mathbf{F}(t)} = \dim M,$$

where $T_{\mathbf{F}(t)}M$ is the vector space tangent to $M$ at the point $\mathbf{F}(t) \in M$ and it was denoted $\dot{f} = df/dt$ for a tensor (or scalar) function $f$ defined on $T$. We define, taking into account Eq.(2.5), the *dissipation function* $\delta$ by

(6.3)
$$\Omega(B_0; t) = -\delta(B_0) dt,$$
$$\delta(B_0; t) = -\dot{E}(B_0; t) - \mathbf{N}(B_0; t) \cdot \dot{\mathbf{F}}(t) + \theta(t) \dot{S}(B_0; t),$$

where $E(B_0; t)$ is the internal energy, $S(B_0; t)$ is the entropy, $\mathbf{N}(B_0; t)$ is the generalized thermodynamic force at the instant $t \in T$. Moreover, it is assumed that although

(6.4) $\quad E(B_0; t) = E(B_0; \lambda(t)), \quad S(B_0; t) = E(B_0; \lambda(t)),$

where $E = E(B_0; \lambda)$ and $S = S(B_0; \lambda)$, $\lambda \in M \times I$, are objective scalars (Section 2), nevertheless the *viscosity effect* of the form

(6.5) $\quad \forall t \in T, \quad \mathbf{N}(B_0; t) = \mathbf{N}(B_0; \lambda(t), \dot{\lambda}(t))$

is, in general, admitted for any thermodynamic process of Eqs.(6.1) and (6.2). Introducing the so-called *net working* at the time $t \in T$ as

(6.6)
$$W(B_0; t) = -\mathbf{N}(B_0; t) \cdot \dot{\mathbf{F}}(t) = V(B_t) \mathbf{T}(B_0; t) \cdot \mathbf{D}(t), \quad \mathbf{T} = \mathbf{T}^{\mathrm{T}},$$
$$\mathbf{N}(B_0; t) = -V(B_0) \hat{\mathbf{T}}(B_0; t) = -V(B_t) \mathbf{T}(B_0; t), \quad B_t = l(\mathbf{F}(t))(B_0),$$

where $\hat{\mathbf{T}}$ and $\mathbf{T}$ are the generalized Piola and Cauchy stress tensors, respectively (Section 2), and we have denoted:

(6.7) $\quad \mathbf{F}^* = (\mathbf{F}^{-1})^{\mathrm{T}}, \quad \mathbf{D} = \frac{1}{2}(\mathbf{L} + \mathbf{L}^{\mathrm{T}}), \quad \mathbf{L} = \dot{\mathbf{F}} \mathbf{F}^{-1},$

we can rewrite the dissipation function in the following form:

(6.8) $\quad \delta = W - \dot{\Psi} - \dot{\theta} S,$

where the relation (2.6) was taken into account.

We generalize the the *dissipation inequality* of Eq.(2.7) assuming that the dissipation function of Eq.(6.3) should fulfill the following condition:

(6.9) $\quad \forall t \in T, \quad \delta(B_0; t) \geq 0,$

equivalent, according to Eq.(6.8), to the so-called *reduced dissipation inequality* [12]:

(6.10) $\quad \dot{\Psi} - W + \dot{\theta} S \leq 0.$

Further on we will consider *homogeneous thermodynamic processes* defined by Eqs.(6.1)-(6.9). A homogeneous thermodynamic process is called *reversible* if the dissipation function vanishes:

(6.11) $$\forall t \in T, \quad \delta(B_0; t) = 0.$$

All other homogeneous thermodynamic processes are called *irreversible*. For example, homogeneous thermodynamic processes are reversible in the case of perfectly thermoelastic nanoclusters (Section 2).

It can be shown (imitating the case of simple thermoelastic materials with the viscosity effect – see e.g. [13]) that if the reduced dissipation inequality is fulfilled, the thermoelastic nanocluster $B_0$ revealing the viscosity effect is *mechanically unconstrained* and the generalized Cauchy stress $\mathbf{T}$ of Eqs.(6.5) and (6.6) is an objective function with respect to variables $(\mathbf{F}, \mathbf{L}) \in GL^+(E^3) \times L(E^3)$, then the stress is independent of the variable $\dot{\theta}$ and can be represented in the following form:

(6.12) $$\mathbf{T}(B_0; \mathbf{F}, \theta, \mathbf{L}) = V(B)^{-1} \left[ \partial_\mathbf{F} \Psi (B_0; \mathbf{F}, \theta) \right] \mathbf{F}^T + \mathbf{T}_D(B_0; \mathbf{F}, \theta, \mathbf{L}),$$

where

(6.13) $$\mathbf{T}_D(B_0; \mathbf{F}, \theta, \mathbf{0}) = \mathbf{0}, \quad \mathbf{T}_D = \mathbf{T}_D^T,$$

and the objectivity condition means that it should be

(6.14) $$\forall \mathbf{Q} \in SO(E^3), \quad \mathbf{T}(B_0; \mathbf{QF}, \theta, \mathbf{QLQ}^T) = \mathbf{QT}(B_0; \mathbf{F}, \theta, \mathbf{L})\mathbf{Q}^T.$$

The dissipation inequality reduces then to the following condition:

(6.15) $$\mathrm{tr}(\mathbf{T}_D \mathbf{D}) \geq 0.$$

Moreover, in this case the formula (2.13) holds [13]. It ought to be stressed that the viscosity properties of small nanomaterial clusters are not satisfactorily recognized. Because of this we will limit ourselves to the case of unconstrained nanoclusters with viscosity effect.

The equivalence of heat and work is represented in the framework of thermodynamic as a relation between the rate of increase of internal energy $E$ due to a mechanical action represented by the net working $W$ of Eq.(6.6) as well as due to a second kind of working, $Q = Q(B_0; t)$, $t \in T$, called the *heating*, which is not identified with anything from mechanics [12]. This relation has the form

(6.16) $$\dot{E} = W + Q,$$

and is called the *first law of thermodynamics*. It follows from Eqs.(6.3), (6.6) and (6.10) that the following condition should be fulfilled:

(6.17) $$\forall t \in T, \quad Q(B_0; t) \leq \theta(t) \dot{S}(B_0; t),$$

what is called the *second law of thermodynamics* (in the Clasius-Planck form, [12]). A (homogeneous) thermodynamic process is considered as a process consistent with the first and second law of thermodynamics. A thermodynamic process is called *adiabatic* if $Q = 0$, *isentropic* if $\dot{S} = 0$, *isothermal* if $\dot{\theta} = 0$ [12]. The case of *perfectly thermoelastic* nanoclusters (Section 2) is characterized by the formula

(6.18) $$Q = \theta \dot{S},$$

and then Eq.(6.16) can be written in the form of the following evolution equation [9]:

(6.19) $$K_\mathbf{F} \dot{\theta} = -\partial_\theta W + Q,$$

where $K_\mathbf{F} = K_\mathbf{F}(B_0; \theta)$ is the *heat capacity* at the constant deformation gradient $\mathbf{F}$ defined as

(6.20) $$K_\mathbf{F} = \partial_\theta E = \theta \partial_\theta S = -\theta \left( \partial^2 \Psi / \partial \theta^2 \right)_\mathbf{F},$$

and Eqs.(2.13), (2.14)-(2.16), (3.6), (6.6), and (6.18) would be taken into account. Note that if the (uniform) temperature $\theta$ attributed to the nanocluster $B_0$ is identified with the temperature of its environment considered as a thermostat (see Section 1), then it seems physically reasonable to restrict ourselves to the case of *isothermal processes* only. It follows from Eq.(6.19) that if $\theta = \theta_0 = \text{const.}$, then

(6.21) $$Q = \partial_\theta W \big|_{\theta = \theta_0}.$$

It follows from Eqs.(6.18) and (6.21) that in this case, the net working is independent of the temperature if and only if this isothermal process is additionally isentropic. Note that since an isothermal and isentropic process is also adiabatic, we are dealing then with the "purely mechanical" case discussed in Section 2.

## 7. Dynamics

Let $B_0 \subset \mathrm{E}^3$ be a nanocluster with the immovable center of mass (identified with its reference configuration) and let $\mathbf{F}: T \to M$, $M \subset GL^+(\mathrm{E}^3)$ be an isothermal (constrained or unconstrained - Section 2) *homogeneous deformation process* at the temperature $\theta \in I$ (see remarks at the very end of Section 6). Let us assume that on the nanocluster act, at each instant $t \in T$, external force fields: the body force field $\mathbf{b}(\mathbf{X}, t)$, $\mathbf{X} \in \text{Int}B_0$, and the surface force field $\mathbf{s}(\mathbf{X}, t)$, $\mathbf{X} \in \partial B_0$, where $\text{Int}B_0$ and $\partial B_0$ denote the interior of the nanocluster and its boundary, respectively. The volumetric kinetic energy $K(B_0; t)$ of the nanocluster (cf. the approximate representation of the mass size-effect by Eq.(2.2)) and the power $P(B_0; t)$ of external forces acting on it are given by:

(7.1)
$$K(B_0; t) = \frac{1}{2} \int_{B_0} \|\mathbf{v}(\mathbf{X}, t)\|^2 dm(\mathbf{X}),$$
$$P(B_0; t) = \int_{B_0} \mathbf{b}(\mathbf{X}, t) \cdot \mathbf{v}(\mathbf{X}, t) dV(\mathbf{X}) + \int_{\partial B_0} \mathbf{s}(\mathbf{X}, t) \cdot \mathbf{v}(\mathbf{X}, t) dF(\mathbf{X}),$$

where it was denoted (see Eq.(2.1), (6.1)) and (6.2)):

(7.2)
$$\mathbf{v}(\mathbf{X}, t) = \frac{d}{dt} l(\mathbf{F}(t))(\mathbf{X}) = \dot{\mathbf{F}}(t)\mathbf{X}, \quad \|\mathbf{v}\|^2 = \mathbf{v} \cdot \mathbf{v},$$
$$dm(\mathbf{X}) = \rho_0 dV(\mathbf{X}), \quad \rho_0 = m / V(B_0).$$

Since the volumetric net working $W(B_0; t)$ of Eq.(6.6) has in an inertial frame reference the following representation [13]:

(7.3)
$$W(B_0; t) = P(B_0; t) - \dot{K}(B_0; t),$$

we obtain, taking into account Eqs.(6.6) and (7.1)-(7.3), that at each instant $t \in T$ and for arbitrary $\dot{\mathbf{F}}(t) \in T_{\mathbf{F}(t)}M$, the following relation holds:

(7.4)
$$\left[\hat{\mathbf{M}}_{\text{ext}}(B_0; t)^T - \ddot{\mathbf{F}}(t)\mathbf{J}(B_0) + \mathbf{N}(B_0; t)\right] \cdot \dot{\mathbf{F}}(t) = 0,$$

where $\hat{\mathbf{M}}_{\text{ext}}(B_0; t)$ denotes the dipole moment of external forces referred to the reference configuration $B_0$ (identified with the nanocluster itself):

(7.5)
$$\hat{\mathbf{M}}_{\text{ext}}(B_0; t) = \int_{B_0} \mathbf{X} \otimes \mathbf{b}(\mathbf{X}, t) dV(\mathbf{X}) + \int_{\partial B_0} \mathbf{X} \otimes \mathbf{s}(\mathbf{X}, t) dF(\mathbf{X})$$

and, according to the condition of immobility of the mass center, the total external force acting on the nanocluster $B_0$ should vanish at each instant $t \in T$:

(7.6)
$$\int_{B_0} \mathbf{b}(\mathbf{X}, t) dV(\mathbf{X}) + \int_{\partial B_0} \mathbf{s}(\mathbf{X}, t) dF(\mathbf{X}) = \mathbf{0}.$$

The generalized thermodynamic force $\mathbf{N}(B_0; t)$ is defined by Eqs.(6.5), (6.6) and additionally by Eqs.(6.12)-(6.15) for unconstrained thermoelastic nanoclusters with viscosity effects, or additionally by Eqs.(2.14)-(2.16) in the case of perfectly thermoelastic nanoclusters with constraints. $\mathbf{J}(B_0)$ is the *tensor of inertia* of the nanocluster $B_0$ determined with respect to its mass center $\mathbf{X} = \mathbf{0}$:

$$\text{(7.7)} \qquad \mathbf{J}(B_0) = \int_{B_0} \mathbf{X} \otimes \mathbf{X} dm(\mathbf{X}).$$

Note that the tensor of inertia is defined in the traditional nomenclature (e.g. in the rigid body mechanics) as $\mathbf{I}(B_0) = \text{tr}\mathbf{J}(B_0)\mathbf{1} - \mathbf{J}(B_0)$ and the tensor $\mathbf{J}(B_0)$ is also called the Euler tensor (of inertia) with respect to the point $\mathbf{X} = \mathbf{0}$ [13].

The following differential equation is a sufficient condition for the validity of Eq.(7.4):

$$\text{(7.8)} \qquad \mathbf{J}(B_0)\ddot{\mathbf{F}}(t)^{\text{T}} = \mathbf{N}(B_0; t)^{\text{T}} + \hat{\mathbf{M}}_{\text{ext}}(B_0; t),$$

or equivalently:

$$\text{(7.9)} \qquad \mathbf{F}(t)\mathbf{J}(B_0)\ddot{\mathbf{F}}(t)^{\text{T}} = \mathbf{M}_{\text{int}}(B_0; t) + \mathbf{F}(t)\hat{\mathbf{M}}_{\text{ext}}(B_0; t),$$

where Eqs.(3.9)-(3.11) with $\mathbf{T} = \mathbf{T}(B_0; t)$ of Eq.(6.6) in place of $\mathbf{T} = \mathbf{T}(B_0; \mathbf{F}, \theta)$ were taken into account. $\mathbf{M}_{\text{int}}(B_0; t)$ is the dipole moment of internal surface forces referred to the instantaneous (actual) spatial configuration $B_t = l(\mathbf{F}(t))(B_0)$, $t \in T$, of the nanocluster. In the case of perfectly thermoelastic nanoclusters, the above equation is equivalent to the condition (7.4) and $\mathbf{M}_{\text{int}}(B_0; t) = \mathbf{M}_{\text{int}}(B_0; \lambda(t))$ is given by Eqs.(3.7), (3.11), (6.1) and (6.2). The "purely mechanical" case [20], discussed in Sections 2 and 6, is closed to the original formulation of the dynamics of affinely-rigid bodies [11].

## 8. Final remarks

The experimental discovery of graphene (being a flat monolayer of carbon atoms tightly packed into a two-dimensional honeycomb lattice) and other free-standing two-dimensional atomic crystals (for example, single-layer boron nitride) [35], makes physically sensible to consider affinely-rigid flat bodies. If $\mathbf{n} \in E^3$ is a distinguished unit vector, then the plane $\pi_\mathbf{n} \subset E^3$ normal to $\mathbf{n}$ and such that $\mathbf{0} \in \pi_\mathbf{n}$ can be identified with a degenerate material space of flat nanoclusters with immovable center of mass (cf. Section 2). Then the space $M_\mathbf{n}$ of mechanical configurations of such nanoclusters can be defined in the following way. Let us denote by $M \subset GL^+(E^3)$ the manifold of mechanical configurations describing *inextensibility* of a three-dimensional nanocluster in the **n**-direction (cf. Section 2 and [13]):

$$(8.1) \quad M = \{\mathbf{F} \in GL^+(E^3): h_\mathbf{n}(\mathbf{F}) = 0\},$$
$$h_\mathbf{n}(\mathbf{F}) = \|\mathbf{Fn}\|^2 - 1 = \mathbf{nCn} - 1, \quad \mathbf{C} = \mathbf{F}^T\mathbf{F} = \mathbf{U}^2,$$

where Eq.(3.8) was taken into account. Next, let us consider the Lie group $U_\mathbf{n}$ of *pure extensions* in the plane $\pi_\mathbf{n}$ (i.e. deformations without rotations located in $\pi_\mathbf{n}$):

$$(8.2) \quad U_\mathbf{n} = \{\mathbf{U} = \mathbf{U}^T \in GL^+(E^3): \mathbf{Un} = \mathbf{n}\}.$$

The space $M_\mathbf{n}$ can be now defined, taking into account Eqs.(3.8) and (8.1), as:

$$(8.3) \quad M_\mathbf{n} = \{\mathbf{F} = \mathbf{RU} \in GL^+(E^3): \mathbf{R} \in SO(E^3), \mathbf{U} \in U_\mathbf{n}\} \subset M.$$

It easy to see that the manifold $M_\mathbf{n}$ fulfils the conditions (M1)-(M3) (Section 2).

A single atomic plane is a flat crystal, whereas 100 layers should be considered as a thin film of a three-dimensional material. But how many layers are needed before a structure is regarded as a three-dimensional one ? For example, single-, double- and few- (3 - <10) layer graphene can be considered as as three different types of two-dimensional crystals ("*graphenes*"). Thicker structures should be considered, to all intents and purposes, as thin films of graphite [35]. From the experimental point of view, such a definition is also sensible. The screening length in graphite is $\approx 5\,\overset{\circ}{A}$ (that is, less than two layers in thickness) and, hence, one must differentiate between the surface and the bulk even for films as thin as five layers [35]. We see that the space $M_\mathbf{n}$ of mechanical configurations defined by Eq.(8.3) can be considered as the general space of mechanical configurations of a two-dimensional nanocluster located in the plane $\pi_\mathbf{n}$ as well as this space can be considered as a space describing inextensibility of a very thin (even in the nanometer observation level scale), three-dimensional nanocluster $B_0$ with its crystal structure consisting of a few atomic planes normal to the **n**-direction. In the latter case the width of a nanocluster $B_0$ in the **n**-direction is treated as a (constant) physical parameter (for example, it can be the case of Eqs.(5.20)-(5.22) with $d = \text{const.}$ and $\beta \ll 1$)

Note also that while, in condensed matter physics, the Schrödinger equation is usually sufficient to describe electronic properties of materials, it is not the case of graphene. Graphene is an exception – its charge carriers mimic relativistic particles and more easily and naturally described starting with a (2+1)-dimensional Dirac-type equation (with the Fermi velocity as an effective speed of light) rather than the Schrödinger equation [35]. So, from the thermomechanical (cf. remarks in [1]) as well as the quantum-mechanical point of views, graphenes represent conceptually a new class of nanostructures and offer inroads into the low-dimensional physics.

**References**

1. P.H. BUFFAT, J.-P. BOREL, *Size effect and melting temperature of gold particles*, Phys. Rev. A., **13**, 2287-2297, 1976.

2. L.I. TRUSOW, V.G. GRYANOW, *Higly dispersed systems and nanocrystals*, Nanostruct. Mat., **1**, 251-254, 1992.

3. C. KOCH, *Bulk behavior of nanostructured materials* [in:] Nanostructure Science and Technology, R.W. SIEGEL, E. HU and M.C. ROCO [Eds.], Kluwer Academic Publishers, Dordrecht, 1999.

4. E.L. NAGAEV, *Small metallic particles*, YFN, **162**, 9, 50-124, 1992 [in Russian].

5. M. BRACK, *Metallic clusters and magic numbers*, Sci. Am. [Polish ed.], 2(78), 34-39, 1998.

6. J.M. MONTEJANO-CARRIZALES, J.L. MORÁN-LOPEZ, *Geometrical characteristic of compact nanoclusters*, Nanostruct. Mat., **1**, 397-409, 1992.

7. B.M. SMIRNOV, *Transition cluster-macroscopic system*, JETP, **108**, 1810-1820,1995.

8. T. BACHELES, H.-J. GÜNTHERODT, R. SCHÄFER, *Melting of isolated tin particles*, Phys. Rev. Lett., **85**, 1250-1253, 2000.

9. A. TRZĘSOWSKI, *Nanomaterial clusters as macroscopically small size effect bodies*, (Part I and II), Arch. Mech., **52**, 159-197, 2000.

10. A. TRZĘSOWSKI, *On the dynamics and thermodynamics of small Markov-type material systems*, arXiv.org e-Print Archive, 2008: http://arxiv.org/abs/0805.0944 .
11. J. SŁAWIANOWSKI, *Analytical mechanics*, PWN, Warsaw 1982, [in Polish].
12. C. TRUESDELL, *Rational thermodynamics*, McGraw-Hill, New York 1969.
13. C. TRUSDELL, *A first course in rational continuum mechanics*, John Hopkins University Press, Baltimore 1972.
14. T. GIEBULTOWICZ, *Breathing life into an old model,* Nature, **408**, 6810, 299-301, 2000.
15. R. SCHÄFER, *Melting of isolated tin nanoparticles*, Phys. Rev. Lett., **85**, 6, 1250-1253, 2000.
16. A. TRZĘSOWSKI, *On the quasi-solid state of solids nanoclusters*, J. Tech. Phys., **44**, 4, 385-396, 2003.
17. A.I. GUSEV, *Effects of nanocrystalline state*, YFN, **168**, 1, 55-83, 1998, [in Russian].
18. A.I.M. RAE, *Waves, particles and fullerenes*, Nature, **401**, 6754, 651-653, 1999.
19. K. SATTLER, *$C_{60}$ and beyond: from magic numbers to new materials*, Jpn. J. Appl. Phys., **32**, 1428-1432, 1993.
20. A. TRZĘSOWSKI, *On constrained size-effect bodies*, Arch. Mech., **36**, 2, 185-193, 1984.
21. A. TRZĘSOWSKI, *Tensility and compressibility of axially symmetric nanoclusters, Part I; simplified modelling,* J. Tech. Phys., **45**, 2, 141-153, 2004.
22. J. MORZYMAS, *Applications of the group theory in physics*, PWN, Warsaw 1997, [in Polish].
23. J. A. WOLF, *Spaces of constant curvature*, University of California, Berkeley 1972.
24. F. SPAEPAN, *Five-fold symmetry in liquids*, Nature, **409**, 781-782, 2000.
25. A. BLINOWSKI, A. TRZĘSOWSKI, *Surface energy in liquids and the Hadwiger integral theorem*, Arch. Mech., **33**, 133-146, 1981.
26. A. HADWIGER, *Altes und neues über konvexe Körper*, Birkhaser, Basel 1955.
27. J. BODZIONY. *A characteristic of spatial structure of crystalline materials*, [in:] Geometrical methods in physics and technology, P. KUCHARCZYK [ed.], WNT, Warsaw 1968, [in Polish].

28. A. COTTREL, *The mechanical properties of matter*, John Willey and Sons, New York 1964.

29. J. CHRISTIAN, *Transformations in metals and alloys, Part I*, Pergamon Press, Oxford 1975.

30. L.A. SANTALO, *Integral geometry and geometric probability*, Addison-Wesley, London 1976.

31. G. PÓLYA, G. SZEGÖ, *Isoperimetric inequalities in mathematical physics*, Princepton University Press, Princepton 1954.

32. M. SYSŁO, *Generalized Stokes law for connected solid figures*, [in:] Geometrical methods in physics and technology, P. KUCHARCZYK [ed.], WNT, Warsaw 1968, [in Polish].

33. G.A. KORN, T.M. KORN, *Mathematical handbook*, McGraw-Hill, New York 1968, [Russian translation, 1984].

34. R.W. SIEGEL, *Creating nanophase materials*, Sci. Am. **275**, 42-47, 1996.

35. A.K. GEIM, K.S. NOVOSELOV, *The rise of graphene*, Nature Materials, **6**, 183-191, 2007.